\documentclass[seceq]{iis}

\usepackage{graphics}
\usepackage{color}
\usepackage{amsmath}
\usepackage{setspace}
\usepackage{caption} 

\subjclass{05.20.-y , 02.70.-c , 02.50.Ng , 02.10.De}

\title{Statistical Mechanical formulation and simulation of prime factorization of integers}
\runtitle{Statistical Mechanical formulation and simulation of prime factorization of integers}

\author{%
\name{\surname{Chihiro H Nakajima}}
\CAE{nakajima@stat.phys.kyushu-u.ac.jp}
}

\inst{$^{1}$Department of Physics, Kyushu University, \address{33 Fukuoka 812-8581, Japan}\\
}

\runauthor{Chihiro H Nakajima}

\abst{
We propose a new approach to solve the problem of the prime factorization, formulating the problem as a ground state searching problem of statistical mechanics Hamiltonian.
This formulation is expected to give a new insight to this problem.
Especially in the context of computational complexity, one would expect to yield the information which leads to determination of the typical case computational complexity of the factorizing process.
With this perspective, first we perform simulation with replica exchange Monte Carlo method.
We investigated the first passage time that the correct form of prime factorization is found and observed the behavior which seems to indicate exponential computational hardness.
As a secondary purpose, we also expected that this method may become a new efficient algorithm to solve the factorization problem. But for now, our method seems to be not efficient comparing to the existing method; number field sieve.
}

\kword{\kw{statistical mechanics}, \kw{prime factorization}, \kw{extended ensemble Monte Carlo method}, \kw{computational complexity}}

\begin{document}
\maketitle

\section{Introduction}
\label{Introduction}
Prime factorization problem is one of relatively few problems called NP-intermediate. 
While no polynomial algorithm has been found, this problem is not considered to be NP-complete\cite{GJ} because of following two facts;
\begin{enumerate}
\item[(i)] The number field sieve method is already known as an relatively efficient algorithm that achieves $O(\exp(n^{\frac{1}{2}}))$ computational time, where $n$ is the logarithm of the composite number which is to be factorized\cite{LL}.
\item[(ii)] As an algorithm for quantum computation, Shor's algorithm is already known to be able to solve in polynomial time\cite{Shor}.
This fact is one of the reasons for that the prime factorization problem is expected to be not NP problem. 
\end{enumerate}
In this proceeding, we propose a statistical mechanical formulation of this problem and numerical analysis of them with Monte Carlo algorithm.

Some investigations in the field between statistical mechanics and computational science have comprehensively explained the threshold behavior of probability of the existence of solutions and the intractability of probabilistic search algorithm near the threshold, based on the phase transition picture of the structure of the space of solutions\cite{MZKST,MMZ,CM}. At first it is discussed on the $k$-SAT problem.
In these researches, the correspondence of the difference of the computational complexity class between $k=2$ and $k \ge 3$ cases to the difference of the behavior of the phase transition is also included.
Similar approaches are applied to other elemental NP-complete computational problems such as the random vertex covering problem\cite{WH}, and random graph coloring problem\cite{ZK2007}.
And the understanding of the computational hardness in terms of the characteristics of energy landscape and the structure of the solution space have been gradually established\cite{MMZ,KMRTSZ,ZK}.
In the same manner as them, we expect to characterize the difficulty of prime factorization problem, on the search algorithm, by investigating the energy landscape and the structure of low energy states in the state space.

In addition to the above aspect, there are also practical applications of statistical mechanics.
A classical example is simulated annealing \cite{KGV}, which is widely used. In some cases, this algorithm has been utilized not only to find the minimum value of cost function itself, but also to find the structure that gives the minima \cite{MSO}.
And as an another example in practical success, the following fact was recognized in the statistical mechanical analysis of NP-complete problems introduced above; even in the groups of problems which are classified as NP-complete, there are several single instances which can be solved in polynomial time.
And, based on the knowledge of the structure of the solution space, an algorithm which works effectively even near the threshold is also proposed \cite{MPZ}, though not able to entirely overcome NP-complete problems.
Therefore, also in practical point of view, it attracts our attention to studying the prime factorization problem with the form that is tractable by probabilistic algorithms which is now conventionally used in the field of statistical mechanics and verifying wheather it can be solved in polynomial time with them. 

\section{Models and Methods}
\subsection{Guidelines for Formulation}
Suppose that an integer $N_o$ is given.
To obtain the prime factorization of $N_o$, we will solve an optimization problem by Markov chain Monte Carlo (MCMC) simulation with the cost function in the phase space.
First, when an integer $N_o$ is located in $2^{n-1}<N_o\leq2^{n}$, the number of its prime divisors is bounded by $n$.
In other words, a number $n$ is defined for each $N_o$ as
\begin{eqnarray}\label{eq:bounding_number}
n=\lceil \log_2 N_o \rceil,
\end{eqnarray}
where $\lceil x \rceil$ is the minimal integer which is larger than $x$.
Let $\{d_i\}$ be the state in the phase space composed of the set of integers $d_i$.
We introduce the two types of formulation in next two sections \ref{sec:whole} and \ref{sec:elemental}.
The maximum number of $i$ is taken as $i=n$ and $i=2$ in section Sec.\ref{sec:whole} and Sec. \ref{sec:elemental} respectively.

Second, the cost function should be designed to favor states in which the elements of $\{d_i\}$ are divisors of $N_o$ and to take its lowest value when the entire set of $\{d_i\}$ is the prime factorization of $N_o$.
Using the information of the residues we can design such cost function.
By definition of the residue, each $\mod(N_o,d_i)$, which is obtained by dividing $N_o$ by each $d_i$, takes the value $0$ only in the case that $d_i$ is a divisor of $N_o$, and becomes positive in the other case.
Instead of $\mod(N_o,d_i)$ itself, we can also adopt following function of $d_i$ and $\mathrm{mod}(N_o,d_i)$,
\begin{eqnarray}
\epsilon_i=\min\bigg(\mathrm{mod}(N_o,d_i) , d_i-\mathrm{mod}(N_o,d_i)\bigg) \ ,
\end{eqnarray}
without loss of such property.

The order of the residue $\mathrm{mod}(N_o,d_i)$ or $\epsilon_i$ is up to that of $O(\exp(n))$.
To formulate a proper statistical mechanical model, we would like to keep the extensiveness of the cost function; Hamiltonian, with respect to $n$.
The reason why we mention about extensiveness is refered in the section \ref{sec:replica_exchange}.
By taking logarithm [see Eq.(\ref{eq:3rd}) below] or using coefficients of $p$-adic expansion $\epsilon_i$ [see Eq.(\ref{eq:pae_epsilon_a})-(\ref{eq:pae_epsilon_c}) below], we can keep such extensiveness.
Especially, with the case of using the $p$-adic expansion coefficient, the extensiveness is naturally guaranteed and its value always becomes integer.
Thus it is particularly convenient in order to calculate the statistical mechanical quantities.

\subsection{Model Hamiltonian}\label{sec:whole}
\begin{figure}[hbt]
\centering
\begin{minipage}{7.8cm}
\includegraphics[width=80mm,keepaspectratio,clip]{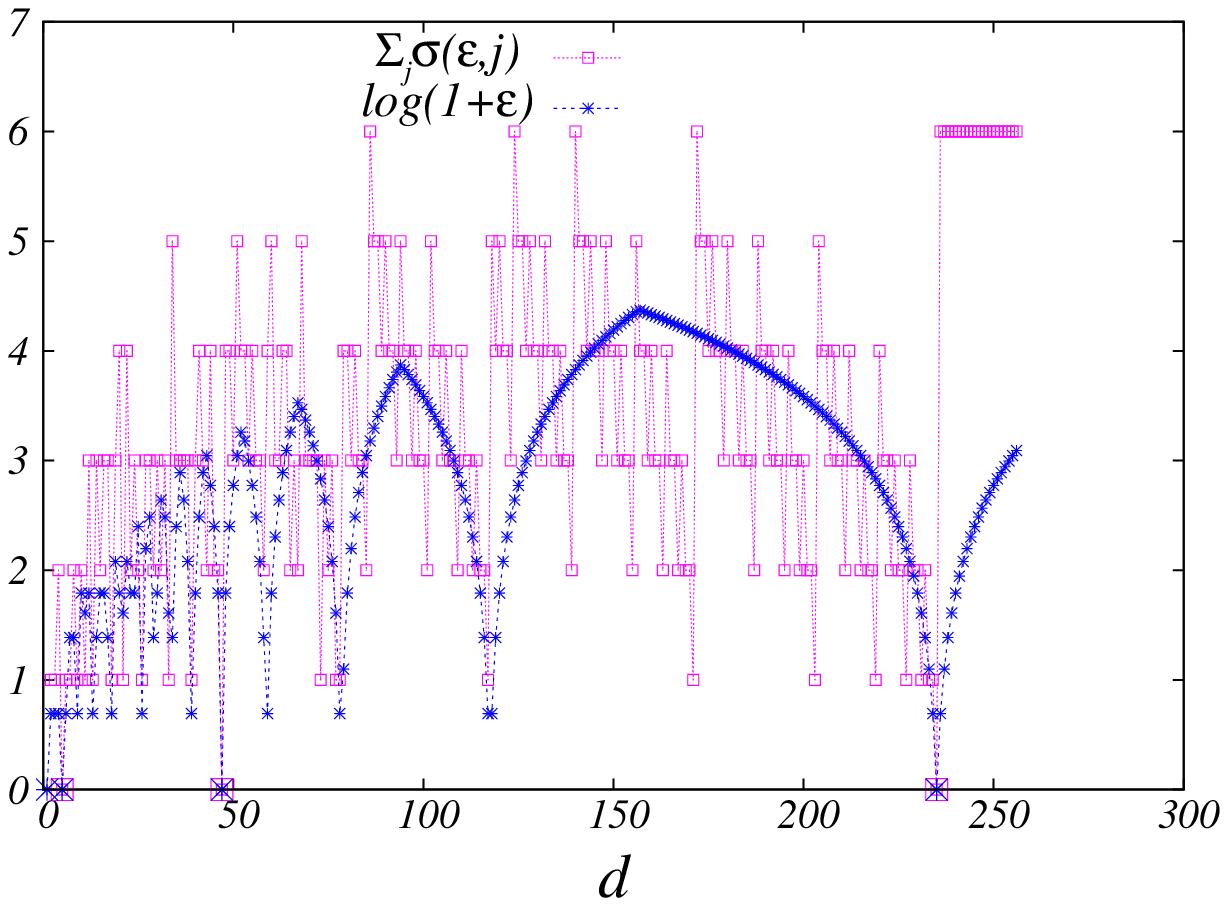}
\caption{The profile of $\log(1+\epsilon)$ introduced in Eq.(\ref{eq:2nd}) and $\sum_{j=1}^{n}\Big(\hat{\sigma}(\epsilon,j)\Big)$ in Eq.(\ref{eq:elem_jyouyo}). Both are in the case of $N_o=235=5 \times 47 < 2^8$. It is indicated that their contribution for each landscape is rough and bounded by 8.}
\label{fig:potential_profile}
\end{minipage}
\hfill
\begin{minipage}{7.8cm}
\includegraphics[width=80mm,keepaspectratio,clip]{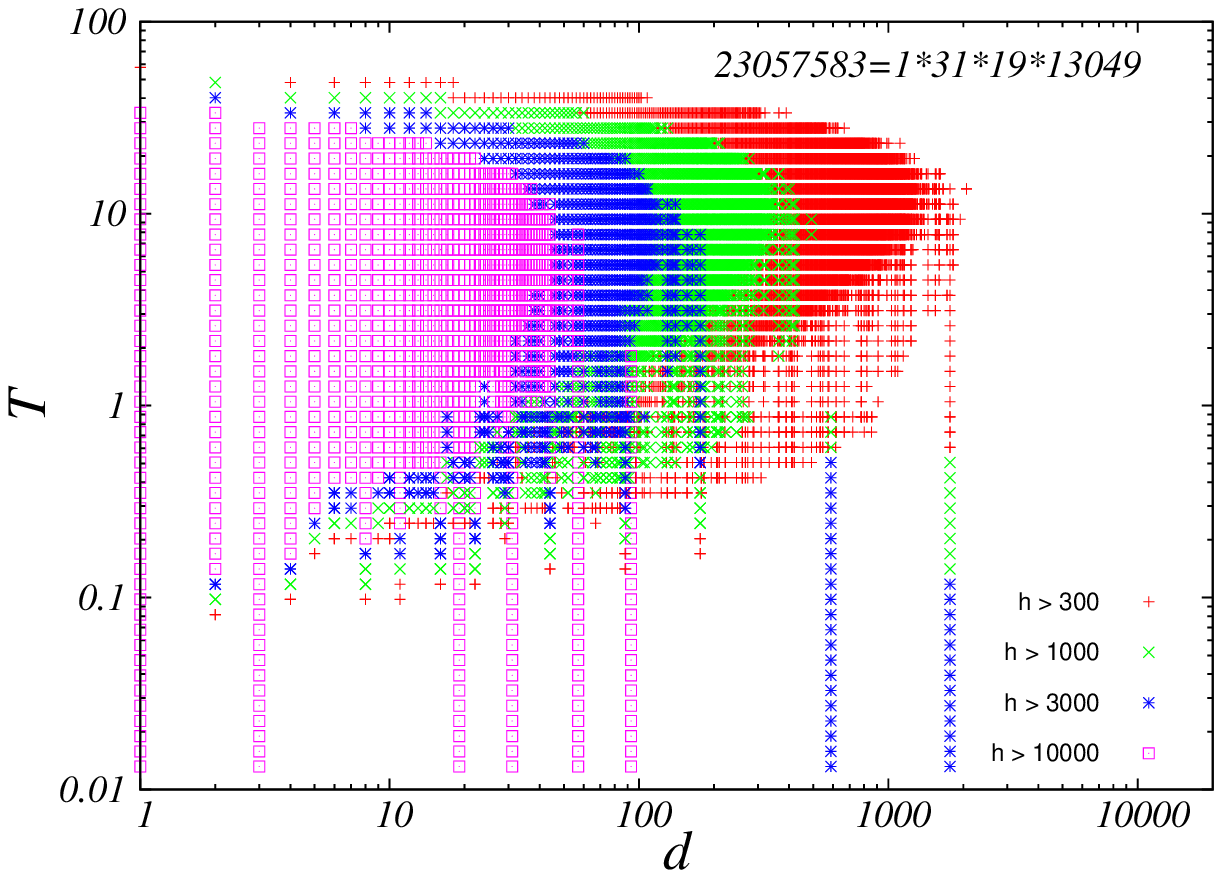}
\caption[8.1cm]{The histogram of the result of the simulation of $H_{\rm{whole}}$ with replica exchange Monte Carlo method in the case of $N_o=23057583$. The value of $\gamma$ is taken to be $0$ in this result. The color red, green, blue, and purple represent the region where the number of hit $h$ is more than $300,1000,3000$ and $10000$ times respectively. At sufficiently low temperature, only the case which $d_i$ is a divisor of $N_o$ is sampled.}
\label{fig:schematic_histogram}
\end{minipage}
\end{figure}
The cost function is mainly composed of two contributions $H_1$ and $H_2$.
$H_1$ comes from the residue terms and $H_2$ does from the difference of the product $\prod_{i}d_i$ from $N_o$.
When representing the prime factorization form by the entire set of $d_i$ we set $i=n$ and prepare $n$ integers $\{d_i\}=\{d_1,\cdots d_n\}$.
Each $d_i$ takes the value $d_i \in \{1, \cdots, 2^{n}\}$.

Thus the detail of the resulting Hamiltonian $H_{\rm{whole}}(\{d_{i}\})$ is shown as,
\begin{eqnarray}
&&  H_{\rm{whole}}(\{d_{i}\} \ | \ N_o)=H_1 + H_2 -\gamma M,  \qquad \left( \ \gamma \geq 0 \ \right) \ \label{eq:whole}\\
&& H_1=\sum_{i=1}^{n}\log(1+\epsilon_i) \label{eq:2nd} \ , \\
&& H_2= \frac{1}{n^2}\bigg(\log N_o-\sum_{i=1}^{n}\log (d_i)\bigg)^2 \label{eq:3rd},
\end{eqnarray}
where $M$ is the number of integers included in $\{d_i\}$ which is larger than $1$. 
$H_1$ takes the value $0$ when all $d_i$ become any divisors of $N_o$.
This term works to each $d_i$ locally.
On the other hand, $H_2$ takes the value $0$ when the product of all $d_i$ is equal to $N_o$.
This term works globally to prohibit the case that all $d_i$ takes the value $1$.
In simulations with sufficiently low temperature, only the states that the full set of $\{d_i\}$ becomes the factorization of $N_o$ are expected to realize [see Fig. 2], because of the term $H_1+H_2$.
And eventually the prime factorization is realized as the ground states of the Hamiltonian (\ref{eq:whole}).

We introduced the $\gamma$ term to treat $N_o$ which can be decomposed to many prime numbers.
But following in this paper, we mainly treat $N_o$ which is composed of only two prime numbers.
Therefore $\gamma$ is set as $0$ in the following of this paper.

\subsection{Another model}\label{sec:elemental}
To formulate the problem of prime factorization as ground state searching, there is some arbitrariness in designing of Hamiltonian.
In practice, it is also efficient to decompose $N_o$ into two divisors recursively and apply the primality test.
In this procedure, the Hamiltonian of factorization of $N_o$ into $d_1 , d_2$ can be written as follows,
\begin{eqnarray}
&&H_{\rm{elem}}(\{d_i\} \ | \ N_o) = H_1+H_2, \label{eq:elem}\\
&&H_1 = \sum_{j=1}^{n}\Big(\hat{\sigma}(\epsilon_1,j)+\hat{\sigma}(\epsilon_2,j)\Big) \label{eq:elem_jyouyo}, \\
&&H_2 = \sum_{k=1}^{n}\hat{\sigma}(|d_1d_2-N_o| \ , \ k) ,
\end{eqnarray}
where the function $\hat{\sigma}( \ , \ )$ represents the coefficient of the $p$-adic expansions of the variables. Here, for an integer $I$, the $j$-th coefficient of $p$-adic expansion $\hat{\sigma}(I,j)$ is defined as
\begin{eqnarray}
I &=&\sum_{j}\hat{\sigma}(I,j)p^{j-1} \label{eq:pae_epsilon_a},
\end{eqnarray}
and each resulting coefficient takes the value $\sigma$ which is in
\begin{eqnarray}
\sigma &\in& \{0,\cdots,p-1\}. \label{eq:pae_epsilon_c}
\end{eqnarray}
The ground states of this Hamiltonian does not become the prime factorization of $N_o$ unless it is the composite number of two prime numbers.
But we can investigate the elemental process of the factorization with this Hamiltonian.

\subsection{Searching with Replica exchange Monte Carlo Method}\label{sec:replica_exchange}
Replica exchange (or some other extended ensemble) Monte Carlo methods are applied to several optimization or constraint satisfaction problems branching from spin glass, including NP-complete problems, and powerful tool for estimating expectation values with less systematic errors, finding the optimal solution, computing  entropy or free energy, counting the number of solutions of these models. In this paper we apply this method to the Hamiltonians formulated as Eq.(\ref{eq:whole}) and Eq.(\ref{eq:elem}).

Changing $d_i$ randomly with each Monte Carlo step, we search the prime factorization of $N_o$ by optimizing the state $H_{\rm{whole}}(\{d_i\}|N_o)$ or $H_{\rm{elem}}(\{d_i\}|N_o)$ with certain condition. 
We have implemented the rule of the transitions in the phase space by representing each $d_i$ with $p$-adic expansion similar to Eq.(\ref{eq:pae_epsilon_a})-(\ref{eq:pae_epsilon_c}) in the cases of the cost function. 
But in this case we adopted the following modulated form of the expansion,
\begin{eqnarray}
d_i&=&1+\sum_{j}\hat{\sigma}(d_i-1,j)q^{j-1}, \\
\sigma &\in& \{0,\cdots,q-1\},
\end{eqnarray}
so that each $d_i$ does not take the value $0$.
We note $q$ instead of $p$ to avoid the confusion.
Thus the phase space is divided-and-conquered by Potts (Ising) like variables $\sigma$. 
Transitions in the phase space are performed by shifting (or flipping) the value of each $\sigma$.
For example, when we adopt $q=2$, the above two Hamiltonians $H_{\rm{whole}}$ and $H_{\rm{elem}}$ are described with $n^2$ and $2n$ degrees of freedom respectively.

In the potential energy landscape of $H_{\rm{whole}}$ or $H_{\rm{elem}}$, like that of spin glass models, there are several local minima [see Fig.\ref{fig:potential_profile}].
Even with probabilistic sampling, in ordinary way,  the random walker can easily become trapped in each minima.
To avoid these trap, it requires iterative heating and annealing of the system.
Thus we perform simulations with several temperature in parallel with exchanging\cite{NH} the correspondence of each walker and temperature.
When a replica is in temperature $T_1$, the microscopic state $\Gamma$ is sampled with stational probability $P(\Gamma,T_1)$.
With certain condition of transition rate, we can keep the canonical stationary distribution in each temperature.

The keystone that enables us to achieve efficient sampling is the exchange between two replicas respectively connected to temperatures which are adjacent each other.
The exchange ratio of two adjacent replicas with energies $E_A$ and $E_B$ is given by $\exp\{(\beta_1-\beta_2)(E_A-E_B)\}$.
As long as the energy function is extensive to the system size $N$, it is known that the number of replicas needed to keep sufficient exchange ration scales with $N^{1/2}$\cite{NH}.
When we adopt the non-extensive energy function $H$, for example $H=|N_o-d_1 d_2|$, we will require significantly and inefficiently larger increase of the number of replicas to keep sufficient exchange ratio.
In statistical mechanical point of view, strictly speaking, there are not necessarily a need to be provided extensiveness of the energy.
In fact, the interesting property of the long range spin glass models with power-law dependence of the energy on the system size are recently researched\cite{BN,BT} and the numerical simulation with replica exchange Monte Carlo method for such systems are also performed\cite{WY}.
However, especially in the case of prime factorization problem, if we attempt to modelize the problem with the way without interposing a complicated function (such as logarithm or p-adic coefficient), the energy function comes to have exponential dependence on its system size as mentioned above.

\section{Result}
\subsection{Behavior of Computational Cost}

We numerically observed the first passage time $\tau_{first}$, the Monte Carlo step that the walker of the MCMC simulation visits the state of the correct factorization for the first time, and its dependence on the system size $n$ both for $H_{\rm{whole}}$ and $H_{\rm{elem}}$.
Various samples of $N_o$ are generated by multiplying two prime numbers which are randomly choosed but moderately close to each other. 

\begin{figure}[hbt]
\centering
\includegraphics[width=80mm,keepaspectratio,clip]{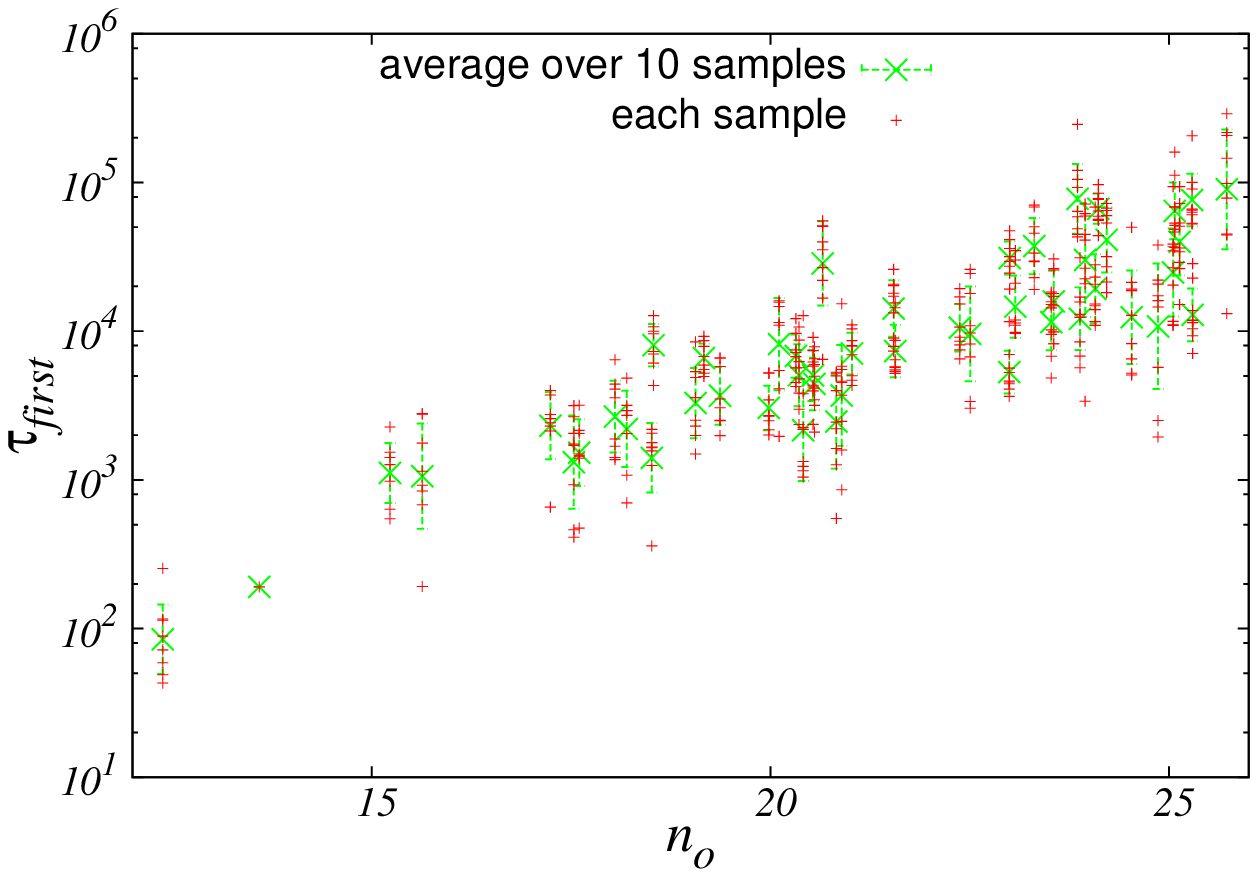}
\caption{The dependecce of $\tau_{first}$ on $n_o=\log_{2}N_o$ with the simulation of $H_{\rm{whole}}$ with semilogrithmic plot. The red points replesents independent $10$ samples for each $n_o=\log_2 N_o$ and the green points represents average over them. The error bar represents the dispersion obtained from $\log\tau_{first}$ of each simulation. It indicates a measure of variation. Trying the numerical fitting to this data with assuming that it has the power law dependence, the exponent is estimated nearly $8$.}
\label{fig:beki_hyouki}
\end{figure}

\begin{figure}[hbt]
\centering
\begin{minipage}{7.8cm}
\includegraphics[width=80mm,keepaspectratio,clip]{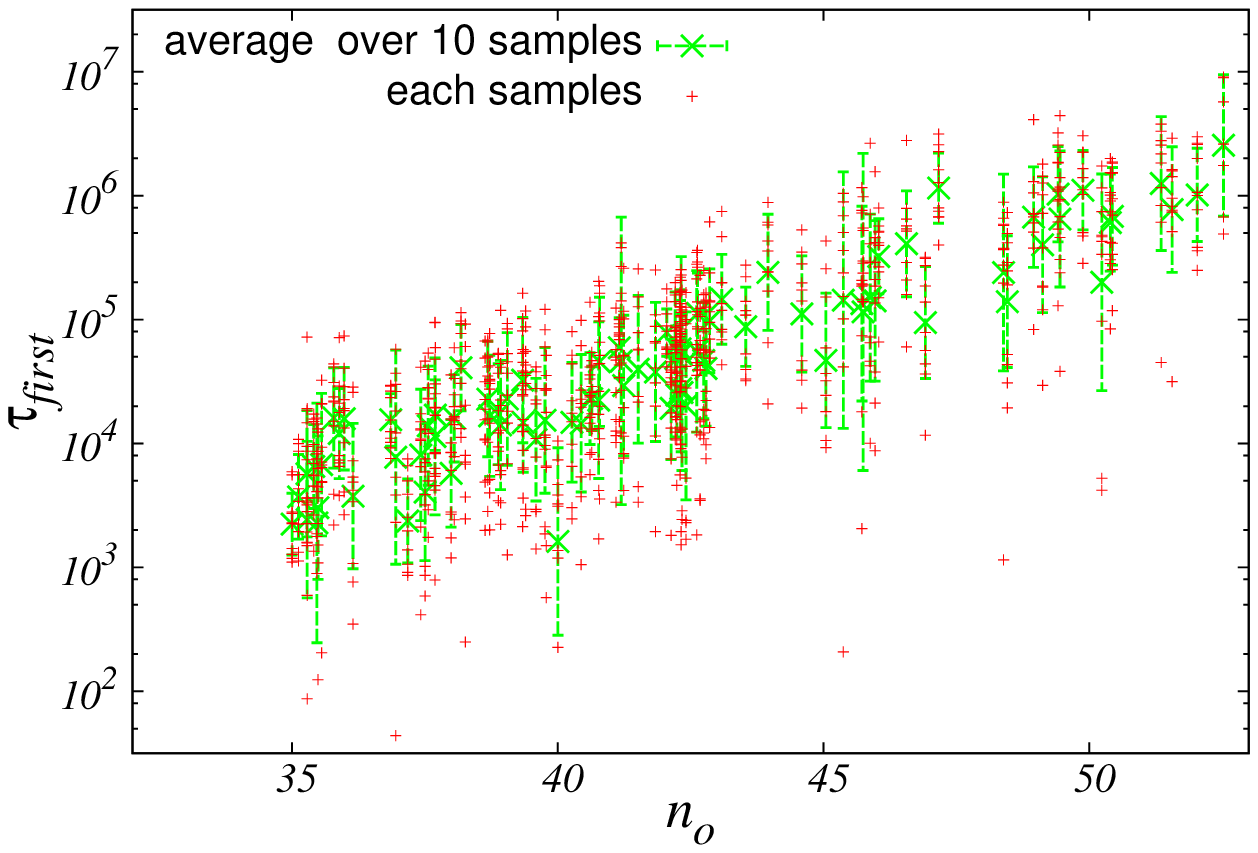}
\caption{The dependecce of $\tau_{first}$ on $n_o$ with the simulation of $H_{\rm{elem}}$ with $q=2$. The means of red points, green points and error bars are the same as Fig.\ref{fig:beki_hyouki} respectively. Both the average and the dispersion are calculated with the same manner.}
\label{fig:p_adic_fpt_p_st2}
\end{minipage}
\hfill
\begin{minipage}{7.8cm}
\includegraphics[width=80mm,keepaspectratio,clip]{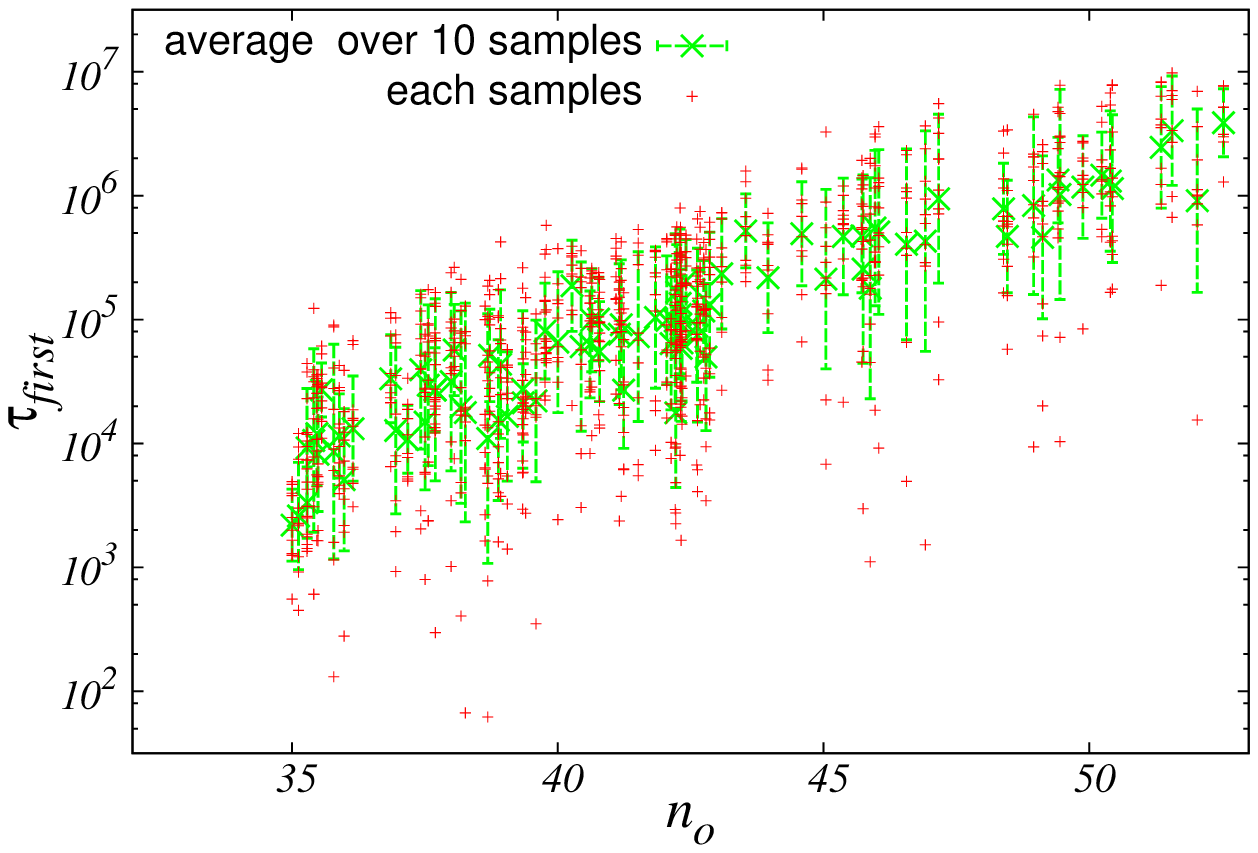}
\caption[8.1cm]{The dependecce of $\tau_{first}$ on $n_o$ with the simulation of $H_{\rm{elem}}$ with $q=3$. The means of red points, green points and error bars are the same as Fig.\ref{fig:beki_hyouki} respectively. Both the average and the dispersion are calculated with the same manner.}
\label{fig:p_adic_fpt_p_st3}
\end{minipage}
\end{figure}

First we start from the explanation of the results of the simulations of  $H_{\rm{whole}}$.
Fig.\ref{fig:beki_hyouki} shows the dependence of $\tau_{first}$ on $n_o=\log_2 N_o$.
Here the results are obtained with $\gamma=0$.
The green points shows the average over $10$ independent samples for each $N_o$.
The average is taken as $\exp\big( \ \overline{\log\tau_{first}} \ \big)$ and the dispersion $\exp\Bigg( \ \sqrt{ \overline{ {\log\tau_{first}}^2 }- \Big( \overline{\log\tau_{first}} \Big)^2 } \ \Bigg)$ indicates a measure of the variation.
Due to the large variation, it is difficult to distinguish between the exponential dependence and power-law dependence with large exponent from these results. 
Even if power-law dependence was actually correct, the exponent is estimated by numerical fitting as nearly $8$. 
In the case of $3$-SAT, in the parameter region which we typically can find a solution in polynomial time, the computational time is nearly proportional to the  system size\cite{MMZ}. 
With respect to that, such a large value in this problem is significantly large. 
At least it is thought to be not practically small value for efficient computing.

Figures \ref{fig:p_adic_fpt_p_st2} and \ref{fig:p_adic_fpt_p_st3} show the results of the case of $H_{\rm{elem}}$.
In this case we performed with two different manner of conquering the phase space, $q=2$ (by Ising variables) and $q=3$ (by Potts variables), with the same $N_o$.
Although both cases have the same values of energy on every microscopic states indicated by $\{d_i\}$, 
the connectivity among each state is different, because of the different conquering manners.
The arrangement of local minima in the phase space is dominated by the connectivity.
Thus the system have different arrangements of local minima with different values of $q$.
In both cases the results are obtained by averaging over $10$ sets of independent simulations for each $N_o$.
These results exhibit similar behavior to the case of $H_{\rm{whole}}$; they are also seen to exhibit exponential dependence on $n_o$ and large (still larger) variation in each sample.
And even with different arrangements of local minima, both $q=2$ and $q=3$ cases exhibit not only qualitatively but also quantitatively similar behavior.
This behavior suggests that low energy local minima appear at statistically (in a single instance of $N_o$) similar density, with little dependence on connectivities among macroscopic states.
This property might be one of the characteristics of the prime factorization problem.

\section{Summary and Discussion}
In this proceeding we have proposed statistical mechanical formulations of the problem of factorization of integers.
We set up two kinds of Hamiltonians and gave rough overviews of the size dependence of their computational cost for optimization by replica exchange Monte Carlo method.
Though the results are not yet sufficient to obtain quantitative conclusion, we observed the behavior which is seemed to indicate that they require exponential computational cost.
However, it should be noted that several questions and issues still remain.

First, the roughly yielded results of the first passage time should be improved by extensive investigation. 
An accurate determination of the probability distribution of the first passage times is left for future work.
As the system size dependence of the distribution directly reflects the computational complexity of this problem, it should be precisely computed.
In the above result about $\tau_{first}$,
the dispersion of $\log\tau_{first}$ calculated from $10$ independent samples is given as a measure of the spread of the distribution function, assuming that their distribution is a Gaussian.
But the tendency of the variation in the figures indicates a possibility that the actual form of the distribution may not be Gaussian.

Second, it should be emphasized that the number field sieve method is already known as an relatively efficient algorithm that achieves $O(\exp(n^{\frac{1}{2}}))$ computational cost.
Considering the fact, it is suggested that the formulation in this proceeding has not yet reached the level of the maximum possible efficiency in the classical computer.
In order to use the above results as any evidence about the computational complexity class with classical probabilistic algorithm, it is thought to be the most reasonable with the case that is confirmed with the most efficient algorithm.
We argue that it is still not excluded out the possibility that we can reach the most efficient level of computation amount by the choice of the way of dividing-and-conquering the phase space, the transition rule, and the temperature points of each replicas\cite{KMS}.
While, it is non-trivial whether we can introduce Markov chain and cost function into the algorithm of the number field sieve method.

As an another subject of future works, we plan to investigate the static quantities of these statistical mechanical models. 
Being trapped in local minima is one of the main cause of a large computational cost on NP-complete problem with local search algorithm has deep relation to its complexity, as revealed with several previous researches\cite{MM}.

In the simulations of this paper, it seemed that the random walker in the phase space had been trapped in several isolated local minima which take the value $1$ or $2$ even in the case that $N_o$ is composed of two prime numbers, when the number of true ground states are order of $O(1)$ in the above Hamiltonians.
As mentioned in the previous section, both cases of $q=2$ and $q=3$ exhibit not only qualitatively but also quantitatively similar behavior, even with different connectivity among microscopic states.
It is thought that the detailed investigation of this characteristic behavior on the landscape can be an effective way in explaining the typical difficulty of prime factorization problem.
In addition, the average behavior of the density of states on energy with various values of $N_o$ would be also useful.
If, suppose that, there is a tendency for the average behavior of density of states that skewed to somewhat certain regions,
one may be able to practically make utilize of it to promote the overall acceleration of the search algorithm  by reweighting the distribution in simulation.
While, if there is no tendency that skewed, that property itself can be an explanation of the specific difficulty of this problem.
By observing the bivariate density of states on the value of energy and the hamming distance from the microscopic state corresponding to the correct factorization, we can enter into the detailed aspect along the above motivation.

It would be effective to observe the density of states focusing on the asymptotic $n$ dependence of the low energy tail.
The characterics of the density of states can also appear in the temperature dependence of specific heat and entropy.
By the detailed analysis of the above quantities, one could gain an understanding of the computational complexity of this problem.



\end{document}